# Automatic microtubule tracking in fluorescence images of cells doped with increasing concentrations of taxol and nocodazole

MarilenaVarrecchia, Joshua Levine, Gabriella Olmo, *Senior Member, IEEE,* Marco Grangetto, *Senior Member, IEEE,* Marta Gai and Ferdinando Di Cunto

***Abstract*—In this paper we explore detection and tracking of astral microtubules, a sub-population of microtubules which only exists during and immediately before mitosis and aids in the spindle orientation by connecting it to the cell cortex. Its analysis can be useful to determine the presence of certain diseases, such as brain pathologies and cancer. The proposed algorithm focuses on overcoming the problems regarding fluorescence microscopy images and microtubule behaviour by using various image processing techniques and is then compared with three existing algorithms, tested on consistent sets of images.**

## I. Introduction

The detection and tracking of objects in digital images is a powerful task, allowing to save time and to improve reliability in the analysis of data in a wide range of areas, such as remote sensing, human-machine interface, biology, medicine. In the biomedical field, image processing techniques are employed for both diagnostic and therapeutic purposes, supporting physicians in their diagnosis and reducing the issues related to the intra and inter-subject variability. This work is focused on the analysis of molecular biology images. In this area, image processing has led to meaningful advances in the early detection, monitoring, and diagnosis accuracy, since it allows to evaluate the cellular and molecular behavior *in vivo*. However, molecules make up complex and dynamic systems generating a big amount of data that require huge processing and a deep knowledge of mathematical models of the molecular structures for their interpretation [1]. As a consequence, even though numerous software-based imaging techniques have been developed to support the biologists in their research, the limited resolution of microscopes and the high dynamicity of the structures of interest impact their accuracy and precision.

Fluorescence microscopy is the main imaging tool used to carry out *in vivo* studies of biological processes, even allowing to check the behavior of single molecules. A major step forward in light microscopy is due to the discovery of a naturally fluorescent protein in living organisms, the *green fluorescent protein* (GFP) [1]. Numerous other markers, with different spectral properties, have been engineered for labelling various types of proteins and cellular structures. This allows biologists to detect specific genes, to evaluate their kinetic parameters, and to evaluate in a quantitative fashion the interactions among molecules [2].

Microtubules are cytoskeleton polymers whose investigation is very significant from the biological point of view. MT are highly dynamic and are involved in different types of cellular functions, such as movements of the inner cell organelles, intracellular transport, and play a crucial role in the cellular division [3]. Modifications in microtubule properties can reveal significant gene mutations, and are responsible of several pathologies, e.g. microcephaly. Nevertheless, the analysis of the dynamics of microtubules in fluorescence microscopy images is challenging, because their dimension (about 20 nm) is below the instruments resolution (about 100 nm) [2]. Hence, one cannot rely on morphological information to detect MTs. Moreover, such images are corrupted by several noise processes (e.g. photon shot noise, background noise, dark current, photobleaching), making their analysis very challenging [1].

Although in the last years several image analysis tools have been proposed for microtubules tracking, because of the complexity of the problem and the lack of a reliable ground truth, the biologists still review the samples manually. It is clear that such a manual job is time-consuming, hardly reproducible, strongly affected by inter and intra-subjects variability. The rest of this paper is organized as follows...

M. Varrecchia and G. Olmo are with the Dept. of Control and Computing Engineering, Politecnico di Torino, Corso Duca degli Abruzzi 24, Torino, Italy. J. Levine and M. Grangetto are with the Computer Science Dept., Corso Svizzera 185, Torino, Italy. M. Gai is with the Department of Molecular Biotechnology and Health Sciences, University of Torino, Via Nizza 52, Torino, Italy. F. Di Cunto is with the Neuroscience Institute Cavalieri Ottolenghi, University of Torino, c.o. Azienda Ospedaliero-Universitaria San Luigi Gonzaga, Regione Gonzole 10, Orbassano, Italy

## II. Microtubules

Along with actin filaments and intermediate filaments, microtubules (MTs) are one of the main building blocks of the cytoskeleton. This latter is involved in several processes, such as spatial organization of the cellular organelles, communication with the extracellular matrix, coordination of the signals aiming to make the cell move and change its shape [4]. In particular, microtubules play a crucial role in the eukaryotic cells, especially for maintaining the genome integrity during the mitosis.

The structural elements of the microtubules are dimers of a globular protein, the tubulin. The dimer is composed of two kinds of polypeptides, $\alpha$-tubulin and $\beta$-tubulin. These dimers polymerize into linear protofilaments; 13 of them, arranged around a hollow core into head-to-tail arrays, make up the microtubule. Because of their structure, the microtubules are polarized elements, and it is possible to recognize two different poles: *plus end* and *minus end*. The first one is characterized by a faster growth speed than the latter, and this impacts on the number of free dimers that bound at the two ends, making the growth happen mainly at the plus end. The microtubules minus-end is linked to the cell centrosome, located near the nucleus in interphase cells, while the plus-ends extend toward the cell cortex.

The dimers can be subject to cycles of assembly (polymerization or growing) and disassembly (depolymerization or shrinking) regulated by the GTP (Guanosine-5'-triphosphate) hydrolysis. Moreover, the polymerization phase is controlled by a third kind of tubulin, the $\gamma$-tubulin [5]. The characteristic microtubule behavior is known as *dynamic instability*, and consists in rapidly alternating phases of growth and shrinkage. The transition from growth to shrinkage is called *catastrophe*, the opposite is called *rescue*. In a third possible state, the *pause*, the microtubule stops growing but does not depolymerize; the factors that regulate this state are still not fully clear. This dynamic behaviour can be altered by a class of proteins selective for microtubules, the MAPs (*microtubule-associated proteins*). These can have different effects on microtubule behavior; some of them are stabilizers because of the capping of the ends, and others act as destabilizers, promoting the depolymerization [5]. In the first case the MAPs make the polymer mass increase, in the second case they reduce it [3]. A third class of proteins, called +TIPs (*Microtubule plus-end tracking proteins*) are selective for the *plus end*, therefore they allow to visually check the growing of the MTs. +TIPs are largely used to evaluate the parameters of interest in living cells, and are also addressed in this paper for MT tracking.

Microtubules play a crucial role in the mitosis process. MT nucleation has a strong impact on the mitotic spindle assembly, hence in the chromosome segregation to two daughter cells during the mitosis [3]. The MTs dynamic instability during the interphase causes the cytoskeleton remodeling in order to properly prepare the cell to the subsequent division and ensure the genome integrity. During this phase, the microtubules move in the surrounding cellular space for few micrometers. This dynamic behavior has an unknown frequency and is a stochastic phenomenon. Three types of microtubules contribute to assembling the mitotic spindle [3]: kinetochore MTs (K-MTs), which link to the chromosomes thanks to a particular protein, the kinetochore; astral MTs (A-MTs), involved in the spindle orientation, which interface with the cellular cortex; and non-kinetochore MTs (nK-MTs), which give stability to the spindle. These give rise to an antiparallel and polarized structure, since the MTs between centrosomes are arranged in array fashion, whereas the ones outside the spindle are arranged in a radial way with the minus-ends directed toward the centrosome and the plus-ends toward the cellular cortex. The effectiveness of the spindle assembly is related to the frequency of catastrophe events and to the growth rate of the microtubules, which is linked to the concentration of free tubulin dimers [3]. When the mitotic spindle is not correctly oriented, abnormal chromosome segregation can occur, and pathologies, due to some gene mutations, can arise. By focusing on the neural cells, it is significant to mention the MCPH (*human primary microcephaly*), a disorder in which the patients are affected by a reduced head circumference involving different degrees of intellectual disability.

The subcellular components and their dynamic behaviour can be analyzed *in vivo* using confocal microscopy. Since the cells and their molecules are transparent in normal conditions, in order to detect a specific particle, markers (or *fluorophores*) having the property to emit light and selective for the proteins to detect are used. The fluorescence microscopy is a powerful tool that allows *in vivo* imaging of molecular structures, but is affected by limitations due to both instrumentation and samples [1]. The result of these limitations is that noisy images are generated, whose analysis is very challenging.

The phenomenon of fluorescence is the molecule capacity to reissue the received radiation. When the radiation is supplied, molecules change their energetic state, which mainly depends on electron configuration, moving from the ground state to a more energetic condition. Fluorescence occurs when the sample comes back to its steady-state, releasing the received energy in form of photons (quantum of energy) [1]. The wavelength of the radiation and the energy are connected by the Planck's law:

$$E = h \cdot v = h \cdot \frac{c}{\lambda}$$

where $E$ is the photon energy, $h$ is the Plank's constant, $v$ is the light frequency related to the speed of light ($c$) and the wavelength of radiation ($\lambda$). Because of heat dissipation, not all excitation energy is emitted in photons; this explains why the wavelength of emission is greater than the excitation one (*Stoke shift*).

To improve the resolution of the system, an immersion layer (e.g. oil) between the lens and the sample is used as a medium; in this way a higher refractive index, and therefore a better resolution, is obtained [1]. The limited resolution (about 100 nm) is not the only factor that affects image quality. Other phenomena that degrade images are related to both instrumentation and samples and can be identified into the low signal-to-noise ratio, the variability of the biological images, the photobleaching, the autofluorescence and phototoxicity that limit the frequency of excitation in *in vivo* cell imaging.

Since the first GFP discovery, different fluorophores have been created thanks to GFP mutations. In this way different genes of interest can be marked and their behavior can be analyzed, also in a quantitative way [2]. The main feature of the targeting molecules is that they must not affect the cellular functions in any way [1]. By focusing on microtubules, the dynamic is usually studied in time-lapse images (2D images over the time) employing tracer build up with tubulin linked covalently to a fluorophore. Only the MTs growth can be monitored using the above-mentioned tracers, which, because of their structure, are selective for the plus ends. This is the reason why these are called *fluorescently-tagged End Binding Proteins* (EB-EGFP). In the experiments involving MT,s mainly the type 1 and type 3 EGFP (EB1-EGFP and EB3-EGFP) are employed. Since the available binding sites for free tubulin decrease exponentially along the microtubule, the fluorescence profile appears in the images as a comet shaped object [6]. The EB-EGFP markers allow to visualize in time-lapse image sequences only the growing MTs, because they bind to plus ends in the assembling phase. Other phenomena such as shrinkages and pauses cannot be directly observed. The cell cultures employed in this work are treated with EB3-tdTomato, a protein belonging to the +TIPs class and therefore selective for the MTs plus ends. The comet detection is the starting point for the microtubules analysis software and it is followed by a tracking step. The existing approaches underlying the current tools will be the subject of the next section.

## III. MT DETECTION AND TRACKING: STATE OF THE ART

The detection and tracking of particles in time-lapse fluorescence sequences have been addressed by many studies, since a quantitative analysis of the reconstructed trajectories yields significant information about functions in living cells [7]. However, nowadays there is no standard protocol to follow because of the extreme variability of the biological processes and the equipment used to acquire the sequences of images. In general, algorithms are divided into four steps [2]:

1) Image data preprocessing to reduce noise levels;
2) Segmentation, that consist in recognizing and sealing off the objects of interest from the background;
3) Particle linkage for tracking the previously identified objects frame by frame;
4) Results analysis for providing quantitative information about microtubules dynamic behaviour.

Besides the fact that fluorescent markers have a size below the optical limits of resolution of the microscopes, the performance of any algorithm in terms of accuracy, robustness and precisionis impaired by the extremely low signal to noise ratio (SNR). Different researches have identified for SNR, specifically in this type of images, a critical value of 4 dB; under this threshold, the reliability of virtually any algorithm quickly impairs [7]. This is the reason why the SNR enhancement is crucial; to this purpose, denoising techniques provide a valid support. Since the predominant noise that corrupts images is not additive, nonlinear filters are used in most of cases [7]. Other problems to tackle are the low contrast of the images due to limited number of fluorescent markers that bind to the plus ends, the autofluorescent background [8], and the fact that microtubules might go out from the focal plane during the experiments. Depending on their features, the existing approaches may be classified into three main categories: probabilistic (bayesian), deterministic, and supervised (i.e. based on machine learning).

### A. *Probabilistic approach*

These tracking approaches are based on the implementation of a filter which is made of two types of models, the dynamic and the measurement model. The first one provides the modelling of the spatial-temporal behavior of the MTs (and more, of the particles of interest), and the second one produces some measurements to include in the first model. The purpose of this filter is to predict the particle positions from a series of measurements. The filtering previously described is often applied through the Kalman filter [8]. It is an estimator of the state of a linear system corrupted by noise and works in optimal conditions if the noise and the error that affects the models are two variables normally distributed with mean zero and not statistically correlated. However, acceptable results can be obtained even if the ideal working conditions are not exactly fulfilled. It is a recursive filter based on two steps: prediction and measurement update. Every time a new measure is detected, its reliability is evaluated by comparing the measured value with the prediction of the system model. Through some matrix operations, a weight is assigned to the measure, and the models of the filter are corrected in order to achieve a better estimation in the next epoch. A more complex application of the Kalman filter is in the Interacting Multiple Model (IMM) filter. It is based on the implementation of a set of Kalman filters to provide a recursive estimation of the model. In this manner more information is retrieved from the history model [9]. The third filter that should be mentioned is the particle filter, it is obtained by the implementation of the Monte Carlo method. This approach is ideal when the models are neither Gaussian nor linear, and ensures a high degree of robustness also when the images are difficult to analyze [8]. A detailed description of the above mentioned filters remains outside the scope of this paper, further information can be found e.g. in [8] [9]. An application of the probabilistic approaches is embedded in the algorithm proposed by P. Roudot et al. [9] Their aim is to construct the particle trajectories in environments with a high density of molecules and subject to rapid motion changes. This scenario leads to an increase in number of false positives, namely objects that have the characteristics to be considered as particles of interest but which should not be detected. The authors tackle the issue following two strategies [9]: Stochastic smoothing or Piecewise-stationary motion modeling (PMM). Their combination is the basis of the piecewise-stationary motion model smoother (PMMS) algorithm. What makes this technique innovative is the analysis of both temporal directions in the time-lapse images for having the maximum quantity of information. The first step plans to detect the particles of interest, therefore for each pixel a Gaussian distribution of the PSF is estimated with a fitting procedure, and a threshold of significance is computed from fitting residuals. In order to estimate the positions of the object centroids, a Gaussian particle intensity model is used to evaluate the location. In particular, the coordinates of the pixel previously labeled as significant and the ones of a local maximum are fitted. The searching of the maxima is carried out in an image filtered with a Laplacian-of-Gaussian filter, while for the fitting procedure a region of interest around the significant pixel is considered [9]. Once the particles have been detected, the next step provides for the tracking and therefore the construction of the trajectories.

The approach proposed by the authors is an update of an existing multiple-particle tracking software made available by Matlab, namely u-track [9]. This is done because, taking into account just the past information, some motion transitions cannot be predicted (e.g. a rapid switch from Brownian to directed motion). Focusing on the microtubules scenario, it can be said that the polymers mainly show a directed motion, rarely a one-dimensional Brownian one. Moreover, an adaptive searching radius for linking particle frame-by-frame is implemented by exploiting both past and future measurements in order to tackle sudden changes. Once the tracks have been identified, they are optimized in a second step which provides for gap closing. It means that the fragmented tracks are connected between them. This process depends on the definition of two parameters, the maximum gap time and the minimum tracks length to take into account. The PMMS algorithm exploits sets of Kalman filters, but each of them is associated to a specific motion regime and the iterations are carried out without mixing the types of motion. The model updating is made in real-time, comparing the new measurement with the other ones available. If a new regime is detected, all Kalman filters are reinitialized with the results of the previous filtering cycle. Since the most recent data on the motion regime are used for the initialization, this represents a suboptimal choice, particularly in the case of rapid changes [9].

The solution proposed by P. Roudot et al. exhibits robustness even when the samples present jerky movements and the frames are acquired with a reduced frame rate, choice that allows to reduce the phototoxicity effects and track poorly labeled molecules. This is the reason why the PMMS estimator is a good decision when the objects to detect are characterized by heterogeneous motions.

### *B. Deterministic approach*

This category of algorithms follow a procedure that can be divided into two macro steps: the localization of particles through enhancing techniques, and the construction of trajectories based on a nearest-neighbor criterion [8]. Generally, the concept of *near* should not be regarded only referred to space, but also to intensity of pixels, shape, direction and other parameters. The majority of approaches implement a search strategy which exploits the intensity of the pixels, in particular the positions of the molecule match with the ones of the intensity of the peaks.

The use of thresholding to detect particle positions is applied in [10]. The approach foresees little *a priori* knowledge of the motion regime and the physic of the molecules; and this makes the procedure less expensive form a computational standpoint [10]. In the initialization phase, all pixels of the sequence are normalized with the min-max scaling; maximum and minimum values of intensity are referred to the entire stack. Then, image enhancing is planned in order to reduce two noise sources: the long-wavelength modulations of background intensity, and the discretization noise that affects the digital images (i.e. photon shot noise) [10]. To this purpose, two low pass filters (LPF) are used, an average filter to remove the background and a Gaussian filter to limit the discretization noise. The first one is based on the assumption that the objects of interest have limited variations compared to the background, therefore a box-car average in a square ROI is implemented. As for the second filter, a normal distribution is assumed for camera noise. The filtered images are the starting point for particle identification. For each of them, the local maximum is evaluated in order to estimate the object positions, but the approach has two main limitations [10]:

- It does not exclude noise;
- It can include false positive particles.

For reducing these issues, a refinement of the positions and the removal of spurious identifications are required. To this aim, the authors fix the particles location, making the assumption that the local maxima detected should not be so far from the real geometric center of the object. An offset is the value that allows to correct the particles coordinates, and it is represented by the distance to the brightness-weighted centroid in the image. The false detections are rejected by assigning a score depending on the intensity profiles. Once the particles positions are estimated in all the frames of the movie, they are linked frame-by-frame to build up the trajectories. To this purpose, a nearest-neighbor criterion is employed. The linking of two particles p and q through two consecutive frames i and j, is based on the minimization of a cost functional, defined as a linear combination of the particle positions, the intensity moments of zero ($m_0$) and second order ($m_2$) [10]:

$$\Phi_{ij} = (\overline{x}_{p_i} - \overline{x}_{q_j})^2 + (\overline{y}_{p_i} - \overline{y}_{q_j})^2 +$$

$$+[m_0(p_i) - m_0(q_j)]^2 + [m_2(p_i) - m_2(q_j)]^2$$

This approach provides good results and is very efficient from a computational point of view [8]. This is the reason why it has been taken as reference in several works over the years, such as the one developed by Sironi et al. and also in this thesis. However, the performance decreases when the images present low quality, and in display environments with a high density of particles.

### C. Combined approach

In some cases, it is possible to devise approaches that mix both the above mentioned techniques. For this category it is worth mentioning the solution proposed by B. Mahemuti et al. [11] that investigate the microtubules dynamic using morphological information for detection and a probabilistic data association (PDA) filter for tracking. The algorithm in [11] starts with the binarization of the images belonging to the movie to analyze, and the removal of all the objects shorter than a threshold depending on the minimum length of microtubules to detect. This parameter impacts the tracking accuracy. Then, the elements identified undergo thinning procedure in order to have microtubules 1 pixel wide [11]. The next step provides for the plus-ends identifications in the binarized images and is based on the assumption that the body of MTs follows the path of head (i.e. plus-end). Taking into account this hypothesis, two consecutive frames will be different just for the positions of the growing ends, and therefore the result of subtracting the two frames is an image containing only the heads of the microtubules. Moreover, since some tracks might cross during the polymers growth, a procedure of objects decomposition is performed in order to isolate single particles forming compound structure. To this purpose, information related to the plus-end positions, the crossing points, and the pixels are exploited. Once the microtubules positions are estimated, the tracks are created with a probabilistic approach considering that the objects could change their shape and topology in different frames of the same video, or that some MTs can suddenly appear or go out from the focal plane. The authors consider the elements into consecutive frames as belonging to the same microtubule if the measured and estimated positions are similar in direction and movement. The PDA filter is based on the Kalman filter, and follows two steps: data association and updating of the tracks. The described technique is characterized by acceptable accuracy in environments with low density of particles, indeed the performance decrease in high density video.

### D. Approach based on machine learning

Machine learning is a powerful tool that allows building up specialized algorithms taking into account the morphology and physic of phenomenon without any a priori knowledge of the particles to analyze. This is possible because the algorithm learns from a collection of data, termed training set. It means that each track in an image is considered as an observation instance and, from the observation of different tracks, it is possible to build a behavior pattern. Once the training procedure is over, the algorithm can evaluate the dynamic of the particles in the movies of interest. However, in the biological field, machine learning is not the most appropriate choice because of the nature of the objects behavior. Indeed, it is a random process, as in the microtubules case, that leads to the lack of a ground truth to use for the training of the algorithm. To build an appropriate data set for the learning procedure, a very comprehensive study of the motion model of the microtubules would be needed. To create such simulator for generating the training set, costs comparable with those of a manual analysis are required. This explains why, even though the huge potential of the machine learning, at present, traditional approaches remain the best choice for molecular images evaluation. In conclusion, it is worth pointing out that regarding the particle detection and tracking, at present, there is not the best algorithm, because all the methods are linked to the experimental dataset. For this reason, the approach to implement should be chosen according to the available dataset.

## IV. Dataset description

For the algorithm testing, a dataset of stacks (i.e. time-lapse image sequences), acquired with a confocal microscope, has been supplied. Cells were treated with two different drugs, nocodazole and taxol, in order to theoretically control their behavior. At high concentrations, these drugs are known to have opposite effects on MTs dynamic. Nocodazole, as a *MTs destabilizer*, promotes MT disassembly, while taxol, as a *MTs stabilizer*, promotes their assembly. Even though their actions on MT dynamics are opposite, the neat effect of these drugs, at high concentrations, is an inhibition of dynamic instability. However, it is worth noticing that the alterations of MTs dynamics are dose-dependent and not completely understood at intermediate concentrations; this is the reason why the cells have been doped with different drug concentrations.

HeLa-K (HeLa Kyoto) cell line, expressing EB3-td Tomato, was chosen to carry out the experiments. HeLa is a cell line class often used for the purpose of scientific research, and is the first human cancer cell line immortalized in tissue culture. They were named HeLa after Henrietta Lacks, a woman that was affected by adenocarcinoma of the cervix, from which cells were extracted with a biopsy [12]. The cell culture was maintained in DMEM-GlutaMAX (Invitrogen) medium supplemented with 10% fetal bovine serum, $100$ U ml$^{-1}$ penicillin, $100$ $\mu$g ml$^{-1}$ streptomycin, $200$ $\mu$g ml$^{-1}$ geneticin (Sigma) and $0.5$ $\mu$g ml$^{-1}$ puromycin.

Interphase cells were treated with increasing concentrations of nocodazole and taxol: 0 nM (control),0.1 nM (taxol only), 1 nM (nocodazole only) 10 nM and 100 nM. After 1 hour incubation, videos of astral microtubules were acquired using a Leica TCS SP5-AOBS 5-channel confocal system, equipped with a 561 nm DPSS laser. During the acquisition, cells were stored in the microscope incubator at 37 °C with $CO_2$ 5%. For each dosage, in both cases, five or six stacks have been acquired and saved in TIFF format. The main characteristics of the image stacks are summarized in Table I.

| Description | Value |
|---|---|
| Frame size | $256 \times 256$ pixels |
| Frame rate | 2 fps |
| Number of frames per stack | 120 |
| Pixel resolution | 64 nm |
| Bit depth | 8 |

TABLE I: Main characteristics of employed images.

### A. Noise description

One of the main types of noise that impairs these images is photon shot noise, caused by the random emission of photons [1]. It becomes relevant when the number of photons is so small that the uncertainty related to the Poisson distribution cannot longer be neglected [13]. Noise intensity is proportional to the square root of the average number of events N (i.e. light intensity). Signal-to-noise ratio, in an image corrupted by shot noise, is $SNR = \frac{N}{\sqrt{N}} = \sqrt{N}$ [13]. Hence, noise levels are stronger when the light source has a reduced intensity and decrease when N is very large [13]. However, source cannot have a high intensity because of *photobleaching*, i.e. the fact that markers lose their capability to fluoresce with time, to an extent related to the light intensity and to the experiment duration. In order to limit photobleaching, the exposure time and the light intensity should be maintained at low levels [1]. It is worth mentioning also the presence of speckle noise; it is a multiplicative noise that degrades images, making them look visually grainy. It becomes relevant when coherent imaging systems are employed, such as laser in confocal microscopy. Noise, in this case, is caused by random interferences between the coherent returns. The effect on grayscale images is an increase of mean intensity in a local area [13].

Another noise source, related to samples, is autofluorescence; it is due to the property of some molecules to naturally fluoresce at wavelengths in the range of visible spectrum, and this emission overlaps with the fluorophore one, making the detection challenging. Other interference sources are background noise, caused by the ambient radiations; dark current, due to the thermal agitation of particles at high temperatures inside the detector, which leads to spontaneous emissions; quantization noise of the digital output; scattering of light, which occurs when the object dimensions are comparable with wavelength size [1]. All the above mentioned interferences lead to a decrease in the overall image contrast and resolution.

It is difficult to isolate microtubules from the surrounding environment since signal levels are comparable with the background ones. It is worth pointing out that particle positions can be detected with an accuracy related to the pinhole detector as follows [2]:

$$\epsilon = \frac{\sigma}{\sqrt{N}}$$

where $\sigma$ is the standard deviation of the instrument point spread function [2], approximated as a normal function, and $N$ is the number of photons detected in the exposure time.

The interference sources could be limited by choosing a small pinhole diameter detector, even if this limits signal intensity. On the other hand, a large diameter will prevent from obtaining the optical confocal effects and other noise sources will be introduced into the videos. Hence, the tradeoff is to set an optimum aperture for the pinhole detector in order to allow signal detection and, at the same time, reject the main noise sources. To improve stack quality and reduce noise levels, all movies have to be managed with image processing techniques.

As mentioned, if the SNR, defined as

$$SNR_d = \frac{\overline{I}_o - \overline{I}_b}{\sigma_b}$$

is below the value of 4 dB, the performance of any algorithm are impaired, as reported in [7]. Table II reports the SNR values evaluated for all the image stacks considered in this paper. It can ben noticed that, even though the available dataset shows an extreme variability in noise levels, most stacks are affected by noise levels below or very close to the critical threshold.

| Nocodazole | | | Taxol | | |
|---|---|---|---|---|---|
| Dose | Stack ID | $SNR_d$ | Dose | Stack ID | $SNR_d$ |
| 0 nM | Ctrl 003 | 4.47 | 0 nM | Resonant | 2.46 |
| | Ctrl 005 | 3.80 | | Ctrl 018 | 3.58 |
| | Ctrl 011 | 6.90 | | Ctrl 024 | 3.52 |
| | Ctrl 013 | 6.23 | | Ctrl 028 | 3.94 |
| | Ctrl 015 | 4.31 | | Ctrl 031 | 3.36 |
| 1 nM | Series 002 | 1.14 | 0.1 nM | Series 003 | 2.28 |
| | Series 005 | 3.52 | | Series 006 | 4.07 |
| | Series 007 | 5.31 | | Series 010 | 3.58 |
| | Series 012 | 4.76 | | Series 015 | 5.29 |
| | Series 015 | 4.76 | | Series 018 | 6.48 |
| | | | | Series 025 | 4.03 |
| 10 nM | Series 003 | 5.31 | 10 nM | Series 006 | 6.47 |
| | Series 005 | 4.62 | | Series 009 | 1.67 |
| | Series 008 | 1.55 | | Series 012 | 3.16 |
| | Series 011 | 3.22 | | Series 015 | 3.60 |
| | Series 014 | 1.96 | | Series 018 | 1.37 |
| 100 nM | Series 002 | 2.58 | 100 nM | Series 002 | 2.30 |
| | Series 004 | 4.01 | | Series 008 | 1.88 |
| | Series 006 | 3.96 | | Series 011 | 3.01 |
| | Series 009 | 2.01 | | Series 016 | 1.24 |
| | Series 011 | 5.31 | | Series 019 | 0.49 |

TABLE II: SNRd values (dB) of the stacks belonging to dataset

## V. The proposed algorithm

The proposed algorithm is divided int three main steps: enhancement, detection and tracking.

### A. Enhancement

Image denoising is performed by means of a LOG-Wiener Transform. First, we apply a logarithmic operator to the image in order to map multiplicative noise into additive one. Then, we process frames with a Wiener filter, based on the working assumptions that noise and signal are not correlated, and the noise process is additive. Finally, the denoised image is obtained by inverse logarithmic transform. The denoising process leads to smooth background, and this facilitates the subsequent particle identification. However, it is worth noticing that the selection of this procedure is driven by heuristic considerations, and the assumptions that the noise is additive and Gaussian are not theoretically guaranteed.

### B. Detection

The algorithm encompasses a calibration phase, which computes a threshold in order to limit false positive detection. To this purpose, we evaluate the global intensity maximum in a user-selected reference image, and set the threshold equal to 35% of that peak. This percentage was chosen after a tuning procedure, in order to detect about a hundred tracks in control stacks. Actual detection is based on a local maxima searching over the frames. Because of the extreme variability of the intensity profile, the search is carried out locally, employing a squared scrolling window (size 7x7 pixels was set following [14]) applied to the enhanced image. Within this window, the local maximum is detected and considered as a plus-end only if its intensity is at least $n$ times the standard deviation of the current frame. The comet positions are refined by centering the squared window on the local maxima previously detected, and recalculating the peak intensity exploring the selected neighbor. This step limits the problem of recognizing as split two objects belonging to the same microtubule. Finally, each peak intensity is compared with the threshold computed in the calibration phase, and it is kept if it overcomes that value, otherwise it is discarded.

### C. Tracking

Tracking is based on the assumption that microtubules exhibit a uniform linear motion. First, the coordinates identified in the previous step are linked in order to build up partial trajectories. To this purpose, plus-end positions are connected frame-by-frame minimizing a cost functional, defined as linear combination of comet coordinates, as proposed in [14]:

$$\Phi_{ij} = (X_{t,i} - X_{t+1,j})^2 + (Y_{t,i} - Y_{t+1,j})^2 \qquad (1)$$

The equation may be regarded as the global energy $\Phi$ of two molecules, $i$ and $j$, between two consecutive frames, $t$ and $t+1$ [14]. Moreover, the maximum displacement allowed between two particles is 7 pixels (i.e. 448 nm). This parameter is critical, since it impacts on the algorithm performance. Once the list of all partial tracks is available, the last step provides for their linking. This avoids considering as different trajectories belonging to the same microtubule because of pause events. Two partial tracks are connected only if the end of the first track and the beginning of the second one have a maximum displacement of 7 pixels and the maximum time lapse between them is 5 frames (i.e. 2.5 s). Finally, as a refinement, tracks are fitted with a second-degree polynomial. Moreover, for the purposes of dynamic evaluation, all tracks shorter than 5 time-points are discarded. They are characterized by Brownian motion, hence they do not meet motion model hypothesis. **enfatizzare differenze rispetto a Sironi!!**

## VI. Results

In our experiments, relevant information on cell behavior is expressed in terms of track velocity, length and lifetime (mean values and standard deviation), and mean number of detected tracks.

As a first general remark, we can notice that, with the parameters selected through calibration, about a hundred tracks are recognized in control stacks. This value is a tradeoff between selectiveness and computational time, and it was selected after a discussion with biologists. Indeed, if this threshold is decreased, more tracks are recognized, and false positive rate is likely to increase. Nevertheless, these trajectories would be anyway discarded downstream of the algorithm, because they do not meet the uniform linear motion model. Moreover, we have verified that the mean and median values (not reported for brevity) exhibit little difference, meaning that outliers have not a significant impact on the algorithm performance.

As a final remark, in both taxol and nocodazole-doped cells, the number of detected trajectories is expected to diminish as the drug concentration increases. In Tables III and IV, it can be seen that this hypothesis is generally confirmed, hence the proposed approach has proven to be conservative.

### *A. Nocodazole-doped cells*

Nocodazole is a MTs destabilizer, meaning that this drug promotes microtubule disassembly and decreases their mass [3]. In Tab. III results are listed for all the four increasing concentrations of nocodazole in terms of velocity, length and lifetime (mean values and standard deviation) and number of detected tracks.

| NOCODAZOLE | | | | | | | | |
|---|---|---|---|---|---|---|---|---|
| Dose | ID | $v_m$ | $\sigma_v$ | $\lambda_m$ | $\sigma_\lambda$ | $\tau_m$ | $\sigma_\tau$ | MTs |
| 0 nM | 003 | 12.24 | 6.18 | 1.25 | 0.70 | 5.71 | 2.59 | 139 |
| | 005 | 15.36 | 7.75 | 1.02 | 0.69 | 4.33 | 2.00 | 113 |
| | 011 | 15.84 | 8.91 | 1.03 | 0.57 | 4.03 | 1.14 | 94 |
| | 013 | 14.98 | 6.80 | 1.33 | 0.74 | 5.09 | 2.62 | 200 |
| | 015 | 16.58 | 9.77 | 1.87 | 1.11 | 5.43 | 2.76 | 258 |
| | **Mean** | **15.00** | **7.88** | **1.30** | **0.76** | **4.91** | **2.22** | **160** |
| 1 nM | 002 | 13.34 | 0.35 | 1.17 | 0.05 | 5.25 | 0.35 | 2 |
| | 005 | 14.00 | 8.04 | 1.25 | 0.59 | 4.57 | 2.01 | 134 |
| | 007 | 16.92 | 9.12 | 1.21 | 0.55 | 3.96 | 1.88 | 85 |
| | 012 | 17.26 | 8.74 | 1.01 | 0.42 | 4.03 | 1.65 | 48 |
| | 015 | 11.97 | 8.73 | 0.65 | 0.46 | 3.75 | 0.82 | 6 |
| | **Mean** | **14.70** | **6.99** | **1.06** | **0.41** | **4.31** | **1.34** | **55** |
| 10 nM | 003 | 18.65 | 8.18 | 1.47 | 0.79 | 4.76 | 2.09 | 190 |
| | 005 | 0 | 0 | 0 | 0 | 0 | 0 | 0 |
| | 008 | 17.87 | 7.30 | 0.96 | 0.39 | 4.22 | 0.91 | 44 |
| | 011 | 0 | 0 | 0 | 0 | 0 | 0 | 0 |
| | 014 | 15.59 | 7.52 | 1.10 | 0.68 | 3.88 | 1.09 | 16 |
| | **Mean** | **17.37** | **7.67** | **1.17** | **0.62** | **4.28** | **1.36** | **83** |
| 100 nM | 002 | 0 | 0 | 0 | 0 | 0 | 0 | 0 |
| | 004 | 0 | 0 | 0 | 0 | 0 | 0 | 0 |
| | 006 | 0 | 0 | 0 | 0 | 0 | 0 | 0 |
| | 090 | 0 | 0 | 0 | 0 | 0 | 0 | 0 |
| | 011 | 12.82 | 7.46 | 1.07 | 0.57 | 4.61 | 1.54 | 137 |
| | **Mean** | **12.82** | **7.46** | **1.07** | **0.57** | **4.61** | **1.54** | **137** |

TABLE III: Nocodazole-doped cells: results

Zero values mean that the stack at hand has not been evaluated. This occurs because noise levels are excessively high, and/or no track can be actually be identified due to the drug effect on MT dynamic behaviour.

From the results reported in Tab. III it can be appreciated that, increasing the drug concentration, fewer stacks are detected; in fact, when the concentration of nocodazole is as high as 100 nM, information can be extracted from a single movie. The velocity trend is also displayed in Fig.1. It can be appreciated that the velocity is basically constant until 1 nM concentration. At 10 nM concentration, the velocity increases significantly, and then decreases again, reaching at 100 nM a value lower than the starting one. This trend was not obvious from the theoretical knowledge of the effects of nocodazole, which is indeed dose-dependent, and may shed some light on the knowledge of the effects of this drug in living cells.

As for mean length of the detected tracks, it decreases at 1 nM, and does not significantly change thereafter. It is worth pointing out that all tracks, on average, are longer than 1 $\mu m$, even if in some cases shorter tracks can be appreciated, e.g. series 015 at 1 nM. This depends on the choice to discard particles not matching the uniform linear motion model.

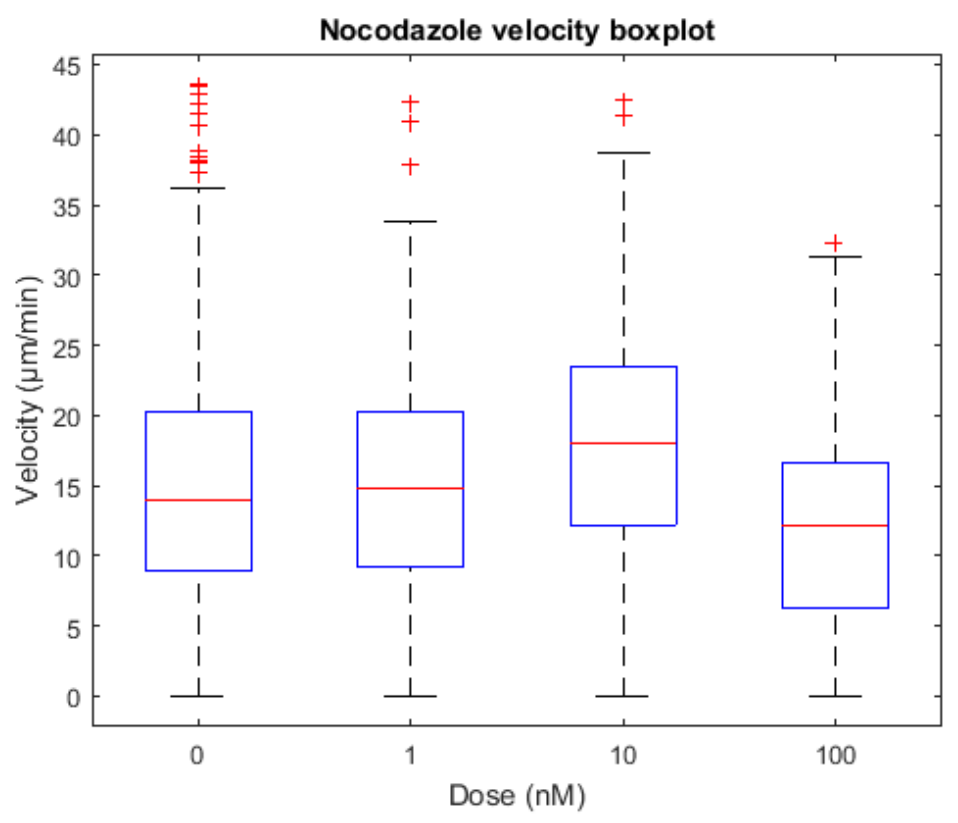

Fig. 1: Velocity trend for different nocodazole concentrations

Lifetime exhibits a trend similar to length, namely it decreases at 1 nM concentration, although a small increase occurs at 100 nM. However, at that concentration only one stack is evaluated, hence this result may be non statistically significant. As already discussed, the number of detected tracks tends to decrease at high concentrations. At 100 nM concentration, in four out of five stacks no track is detected.

Finally, for all the considered parameters, the standard deviation is significant. Actually, Fig. 2 reports the sample distributions of velocity, length and lifetime in control stacks (similar results can be obtained also for other drug concentrations).

It can be noticed that the velocity sample distribution is approximatively normal, whereas Fig. 2(b) and 2(c) exhibit an exponential decay for both length and lifetime. The length distribution exhibits a larger sample standard deviation, if compared with velocity or lifetime; this feature can be regarded as the most critical to estimate.

### *B. Taxol-doped cells*

Taxol is a MTs stabilizer, so it leads to increased polymer mass and suppresses microtubule dynamics [3]. As result, fewer and shorter MTs are expected to be detected, even if the global effects depend on the tested concentration. In Tab. IV the results obtained from taxol-doped samples are reported in terms of velocity, speed and lifetime (mean value and standard deviation) and number of detected comets.

Comparing the nocodazole and taxol results in control stacks (where actually no drug is employed), we can notice that in the second case molecules exhibit higher mean velocity. This cannot be explained in terms of drug effects, since cells have not been doped in either case. This suggests the extreme variability and complexity of the problem, since cell functions are altered not only by drugs, but also by environmental factors (e.g. temperature). Differently to the nocodazole case, in taxol-doped case the algorithm is able to extract information from all the available stacks.

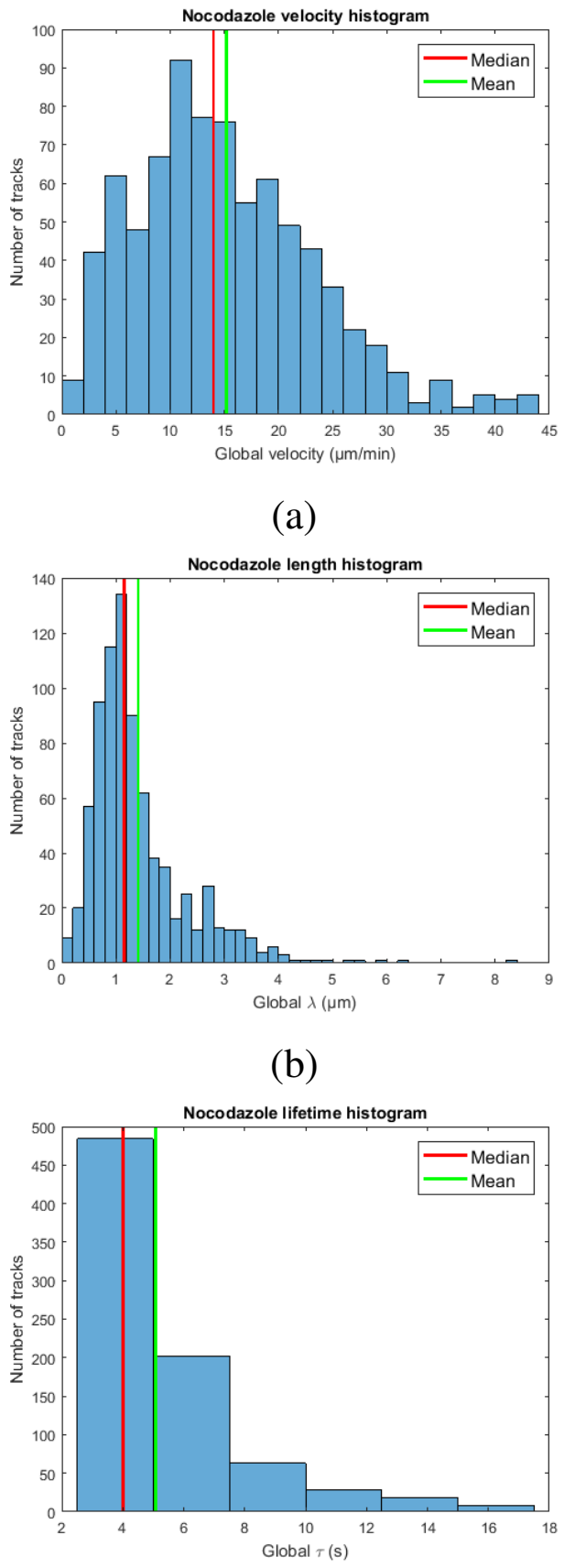

(a)

(b)

(c)

Fig. 2: Nocodazole-doped cells: sample distributions of velocity, speed and lifetime

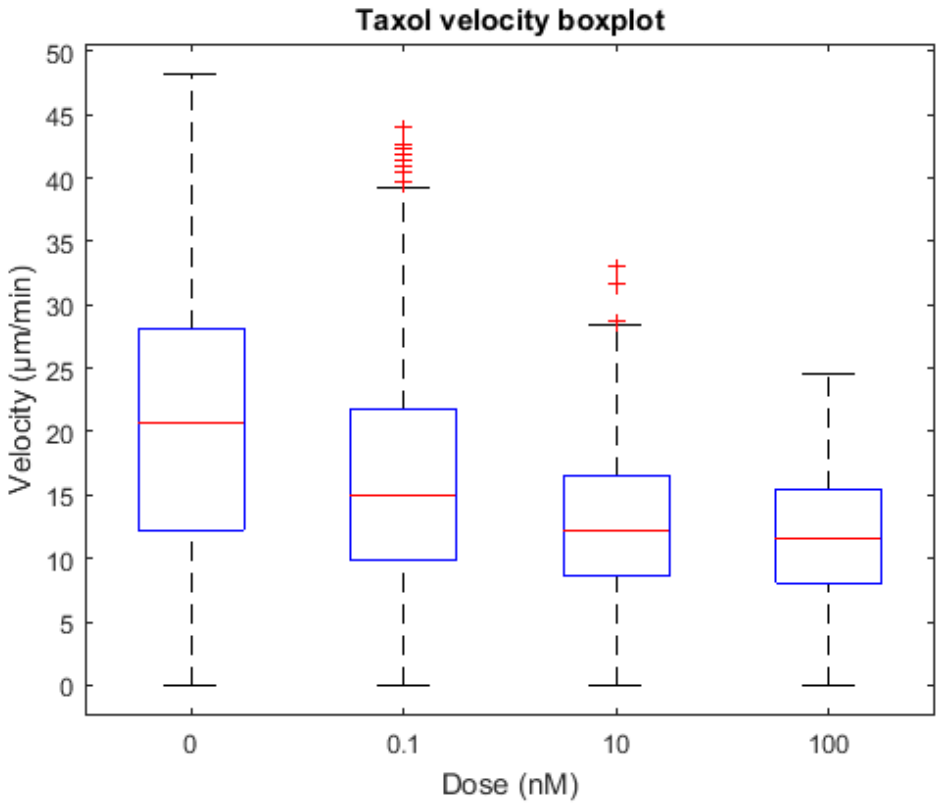

Fig. 3: taxol boxplot

Velocity mean values (also reported in Fig. 3 for the sake of clarity) show a uniformly decreasing trend, and this is coherent with the theoretical knowledge of the taxol effect on MT dynamic behaviour. Moreover, analyzing single stacks, series 011 at 100 nM can be considered as outliers, since they yield anomalous velocity values that strongly bias the global mean.

| TAXOL | | | | | | | | |
|---|---|---|---|---|---|---|---|---|
| Dose | ID | $v_m$ | $\sigma_v$ | $\lambda_m$ | $\sigma_\lambda$ | $\tau_m$ | $\sigma_\tau$ | MTs |
| 0 nM | Res | 21.32 | 9.05 | 1.65 | 1.06 | 4.19 | 1.39 | 84 |
| | 018 | 22.61 | 10.49 | 1.42 | 0.72 | 4.14 | 1.63 | 69 |
| | 024 | 19.01 | 10.55 | 1.24 | 0.79 | 4.32 | 1.54 | 125 |
| | 028 | 21.14 | 9.07 | 1.11 | 0.71 | 3.77 | 0.96 | 11 |
| | 031 | 19.76 | 11.42 | 1.36 | 0.84 | 3.81 | 1.35 | 40 |
| | **Mean** | **20.77** | **10.12** | **1.36** | **0.82** | **4.05** | **1.37** | **65** |
| 0.1 nM | 003 | 15.42 | 8.08 | 1.47 | 0.90 | 4.54 | 2.11 | 312 |
| | 006 | 13.55 | 7.30 | 1.16 | 0.66 | 4.60 | 1.83 | 147 |
| | 010 | 15.47 | 9.03 | 1.59 | 1.66 | 5.01 | 3.14 | 152 |
| | 015 | 15.71 | 8.13 | 1.48 | 0.90 | 4.35 | 1.64 | 283 |
| | 018 | 17.34 | 8.64 | 1.51 | 0.73 | 4.47 | 1.68 | 369 |
| | 025 | 18.00 | 8.87 | 1.46 | 0.88 | 4.80 | 2.08 | 233 |
| | **Mean** | **15.91** | **8.34** | **1.44** | **0.96** | **4.63** | **2.08** | **249** |
| 10 nM | 006 | 13.26 | 5.67 | 1.33 | 0.70 | 5.52 | 2.56 | 125 |
| | 009 | 13.27 | 7.08 | 0.86 | 0.48 | 3.75 | 0.50 | 4 |
| | 012 | 12.09 | 5.77 | 1.38 | 0.81 | 5.19 | 2.83 | 106 |
| | 015 | 14.24 | 6.98 | 1.46 | 0.76 | 4.85 | 1.84 | 88 |
| | 018 | 14.43 | 5.36 | 0.68 | 0.31 | 3.67 | 0.97 | 9 |
| | **Mean** | **13.46** | **6.17** | **1.14** | **0.61** | **4.59** | **1.74** | **66** |
| 100 nM | 002 | 13.61 | 5.78 | 0.92 | 0.55 | 4.27 | 1.41 | 30 |
| | 008 | 14.44 | 3.52 | 1.54 | 0.60 | 6.31 | 1.93 | 8 |
| | 011 | 21.51 | 0.00 | 1.08 | 0.00 | 3.00 | 0.00 | 1 |
| | 016 | 11.51 | 6.18 | 0.99 | 0.56 | 5.25 | 2.62 | 57 |
| | 019 | 7.68 | 0.00 | 0.38 | 0.00 | 3.00 | 0.00 | 1 |
| | **Mean** | **13.75** | **3.10** | **0.98** | **0.34** | **4.36** | **1.19** | **19** |

TABLE IV: Taxol-doped cells: results

The drug effect on the length begins to be apparent at 10 nM, when tracks become shorter, whereas at low concentrations (0 and 0.1 nM) no relevant modification occurs. Finally, lifetime increases already at concentrations as low as 0.1 nM; this can be explained by the freezing effect on the MT dynamic behaviour caused by taxol, that slows down growth without depolymerizing microtubules. As for the number of detected tracks, we can notice that in stacks doped with 0.1 nM taxol more tracks are detected than in control stacks. This can be due to the overall higher intensity of those movies, leading to higher false positive object detection. This is the reason why the detection thresholds should be readapted, making the algorithm more selective in this specific case. However, it is worth pointing out that this only impacts the algorithm performance from a computational standpoint, since the shortest tracks are not considered for subsequent feature evaluation, as already discussed.

Fig. 4 shows the sample distributions of speed, length and lifetime for a taxol concentration of 0.1 nM. The same consideration already made as for Fig. 2 still hold true in this case.

### C. *Statistical data analysis*

In order to assess the statistical reliability of the obtained results, the standard error of the mean (SEM) has been worked out for velocity, length and lifetime. SEM is an indicator of the value variability among different experiments, and it is defined as:

$$SEM = \frac{\sigma}{\sqrt{N}}$$

where $\sigma$ is the standard deviation of the distribution of the parameter at hand, and $N$ is the sample size. In this work, for each drug concentration, the sample standard deviation has been employed, whereas $N$ is the cumulative number of tracks detected in each stack. Table V reports the SEM for the three parameters taken into account and both taxol and nocodazole-doped cells.

It can be noticed that the SEM values are quite low, hence we can conclude that the obtained valued at each concentration, are quite sound from the statistical point of view.

Finally, the three features, namely velocity, length and lifetime, have been estimated independently; however, their mean values are indeed correlated. In the following, we assume that two variables (i.e. velocity and lifetime) as independently estimated, and we compute the third one (i.e. the length) from the mean values of the others. Therefore, length is computed as:

$$\lambda = \frac{v\tau}{c}$$

where $v$ represents velocity (in $\mu$m/min), $\tau$ the lifetime (in s), and c is a conversion factor. In Table VI the obtained length values are compared to the ones estimated by the algorithm.

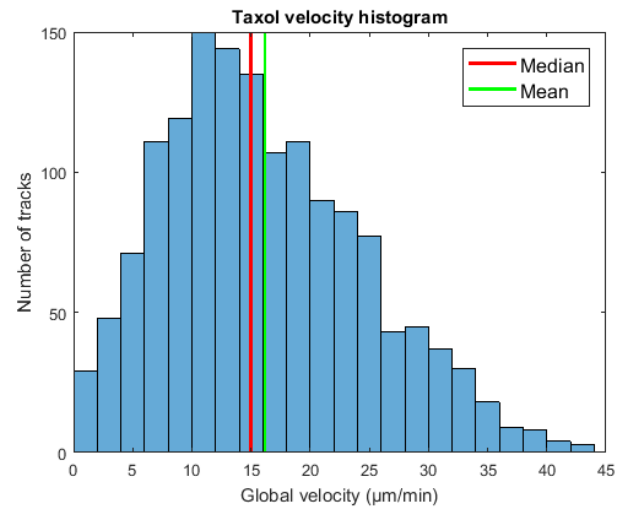

(a)

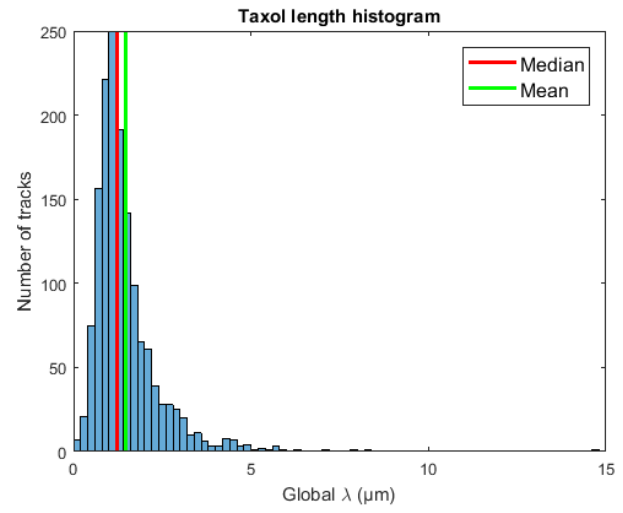

(b)

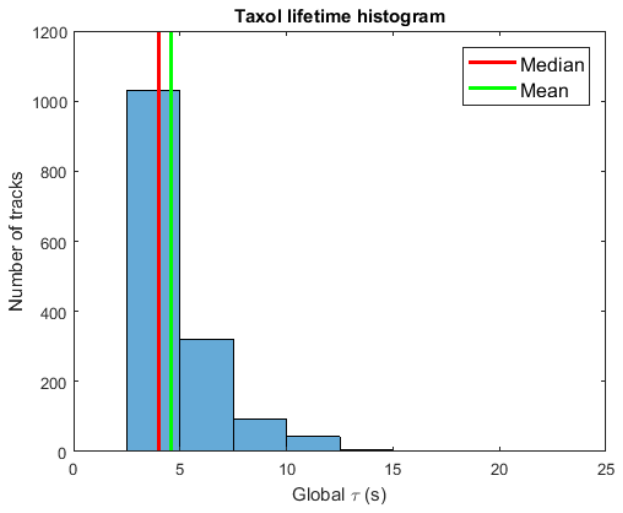

(c)

Fig. 4: Taxol-doped cells: sample distributions of velocity, speed and lifetime

| NOCODAZOLE | | | |
|---|---|---|---|
| Dose | $SEM_v$ | $SEM_\lambda$ | $SEM_\tau$ |
| 0 nM | 0.28 | 0.03 | 0.08 |
| 1 nM | 0.42 | 0.03 | 0.08 |
| 10 nM | 0.48 | 0.04 | 0.09 |
| 100 nM | 0.64 | 0.05 | 0.13 |
| TAXOL | | | |
| Dose | $SEM_v$ | $SEM_\lambda$ | $SEM_\tau$ |
| 0 nM | 0.56 | 0.05 | 0.08 |
| 0.1 nM | 0.22 | 0.02 | 0.05 |
| 10 nM | 0.34 | 0.03 | 0.10 |
| 100 nM | 0.31 | 0.03 | 0.12 |

TABLE V: Standard error of the mean for velocity, speed and lifetime. Taxol and nocodazole-doped cells at different concentrations

It is clear that estimated length values do not significantly differ from those evaluated from velocity and lifetime mean values. This is in favour of a good algorithm reliability.

| NOCODAZOLE | | |
|---|---|---|
| Dose | $\lambda_a$ | $\lambda_e$ |
| 0 nM | 1.30 | 1.23 |
| 1 nM | 1.06 | 1.06 |
| 10 nM | 1.17 | 1.24 |
| 100 nM | 1.07 | 0.99 |
| TAXOL | | |
| Dose | $\lambda_a$ | $\lambda_e$ |
| 0 nM | 1.36 | 1.40 |
| 0.1 nM | 1.44 | 1.23 |
| 10 nM | 1.14 | 1.03 |
| 100 nM | 0.98 | 1.00 |

TABLE VI: Estimated vs. computed length

## D. *Performance comparisons*

As previously discussed, there is no standard protocol for microtubule tracking, because of the extreme variability of the involved biological processes. Hence, a ground truth to refer to is not available. For this reason, in order to assess the algorithm performance and reliability, the results obtained by the approach described in this work (labelled *Algorithm 1*) are compared with those achieved by the algorithm in [15], labelled *Algorithm 2*, which employs the same data sets used in the present work, making the comparison very significant. Finally, our approach is compared with that proposed by Applegate et al. in [16] (*Algorithm 3*) and that described in [14] (labelled *Algorithm 4*). All the comparisons, in terms of average velocity and length and the respective standard deviations, are summarized in Tab. VII for both nocodazole and taxol-doped cell cultures. If data are not available for a given stack, the table reports NA (*not available*).

| NOCODAZOLE | | | | | |
|---|---|---|---|---|---|
| Dose | Algo | $\overline{v}$ | $\sigma_v$ | $\overline{\lambda}$ | $\sigma_\lambda$ |
| 0 nM | 1 | 15.00 | 7.88 | 1.30 | 0.76 |
| | 2 | 16.23 | 13.43 | 1.00 | 1.41 |
| | 3 | 20.57 | 13.00 | 0.63 | 0.52 |
| | 4 | 16.00 | 0.75 | 1.45 | 0.30 |
| 1 nM | 1 | 14.70 | 6.99 | 1.06 | 0.41 |
| | 2 | NA | NA | NA | NA |
| | 3 | NA | NA | NA | NA |
| | 4 | NA | NA | NA | NA |
| 10 nM | 1 | 17.37 | 7.67 | 1.17 | 0.62 |
| | 2 | 19.92 | 15.53 | 1.12 | 1.49 |
| | 3 | 18.56 | 10.20 | 0.61 | 0.53 |
| | 4 | NA | NA | NA | NA |
| 100 nM | 1 | 12.82 | 7.46 | 1.07 | 0.57 |
| | 2 | 11.20 | 10.51 | 0.34 | 0.40 |
| | 3 | 17.33 | 11.60 | 0.44 | 0.32 |
| | 4* | 13.10 | 2.25 | 1.00 | 0.24 |
| TAXOL | | | | | |
| Dose | Algo | $\overline{v}$ | $\sigma_v$ | $\overline{\lambda}$ | $\sigma_\lambda$ |
| 0 nM | 1 | 20.77 | 10.12 | 1.36 | 0.82 |
| | 2 | 22.67 | 17.92 | 1.20 | 1.70 |
| | 3 | 20.44 | 11.10 | 0.67 | 0.58 |
| | 4 | 15.50 | 1.40 | 1.60 | 0.23 |
| 0.1 nM | 1 | 15.91 | 8.34 | 1.44 | 0.96 |
| | 2 | NA | NA | NA | NA |
| | 3 | NA | NA | NA | NA |
| | 4 | NA | NA | NA | NA |
| 10 nM | 1 | 13.46 | 6.17 | 1.14 | 0.61 |
| | 2 | 11.06 | 9.34 | 0.45 | 0.55 |
| | 3 | 15.25 | 11.60 | 0.40 | 0.32 |
| | 4** | 9.00 | 1.40 | 0.95 | 0.15 |
| 100 nM | 1 | 13.75 | 3.10 | 0.98 | 0.34 |
| | 2 | 8.66 | 8.86 | 0.29 | 0.34 |
| | 3 | NA | NA | NA | NA |
| | 4 | 7.00 | 1.00 | 0.40 | 0.15 |

TABLE VII: Comparison among different algorithms. * concentration of 80 nM. ** concentration of 20 nM

If we focus on the comparison between Algorithms 1 and 2, we can appreciate that they yield fairly coherent trends as for mean velocity and length in most cases, with Algorithm 1 generally providing slightly lower values, due to the fact that Algorithm 2 is tuned so as to yield a larger number of detected tracks. We point out the significant difference in average length at 100 nM concentration of both taxol and nocodazole. This is due to the screening process, implemented in the algorithm proposed in this work, which removes all short tracks not matching the uniform linear motion assumption. This also explains why in the second algorithm more tracks are detected than in the first one. The last aspect to be noticed is the different velocity and length variability range in the two mentioned approaches. However, standard deviation difference is sharper for speed case than for the length one. For the sake of clarity, the mean velocity and length provided by Algorithms 1 and 2 are also reported in Fig. 5 for nocodazole-doped cell cultures. We can appreciate that the two algorithms exhibit the same trend for both variables. As already discussed, the only remarkable difference is the length value at 100 nM.

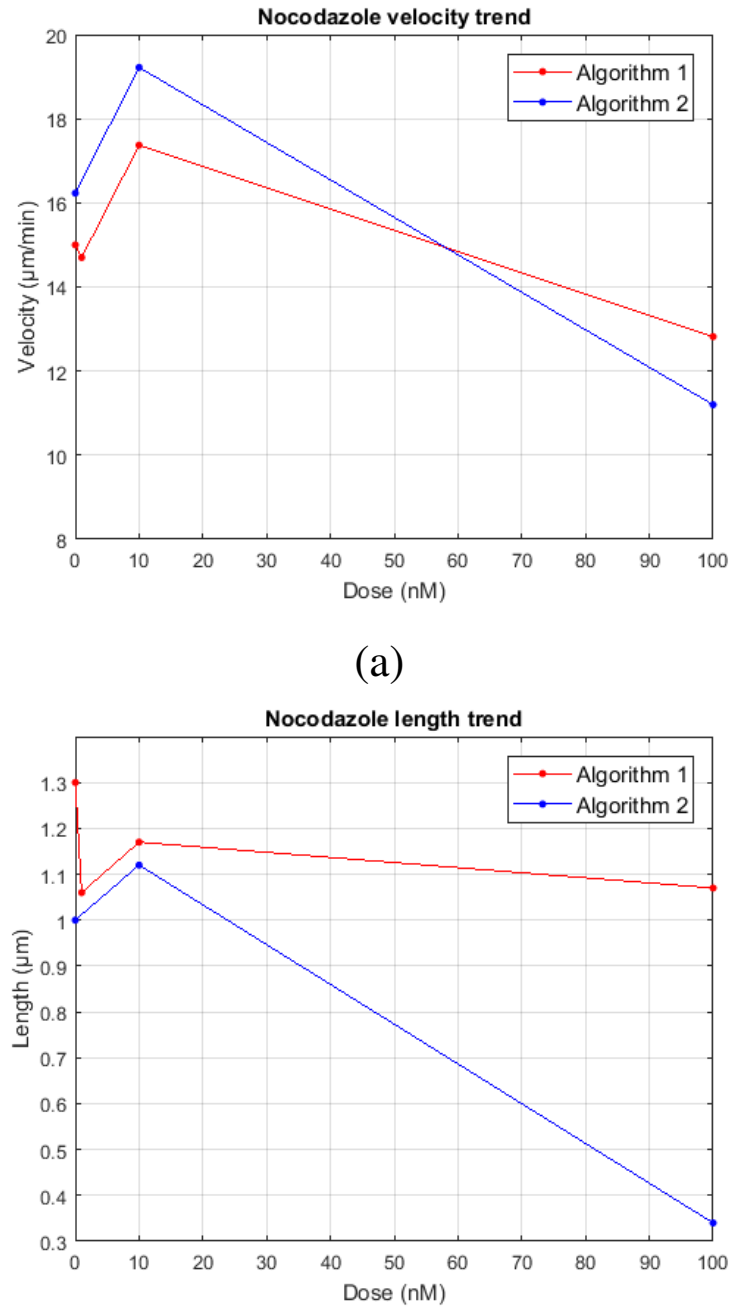

(a) (b)

Fig. 5: Nocodazole-doped cultures: comparison between Algorithms 1 and 2

As for the comparisons with Algorithms 3 and 4, it can be noticed that Algorithm 3 generally yields higher mean velocity and standard deviation values if compared to Algorithm 1, whereas Algorithms 1 and 4 exhibit similar values in control stacks and at 100 nM drug concentrations. Moreover, it is worth noticing that Algorithm 4 yields very low standard deviation, due to an algorithmic choice that suppresses variability to a large extent. As for length, both Algorithm 1 and Algorithm 4 yield nearly the same average values; instead, Algorithm 3 is able detect much shorter tracks. This difference can be justified by the removal of the shorter tracks carried out by both Algorithm 1 and Algorithm 4, but not by Algorithm 3. Moreover, the three algorithms show a comparable length standard deviation. However, it is worth emphasizing that Algorithms 3 and 4, although addressing nocodazole and taxol-doped cells, employ different cell cultures, and this actually make quantitative comparisons not significant. In any case, the velocity and length yielded by the algorithms in the case of nocodazole-doped cell cultures are reported in Fig. 6. We point out the fact that, whereas Algorithms 1 and 2, at intermediate concentrations (i.e. 10 nM), show an increase of MTs dynamicity in terms of velocity (see also Fig. 5), this behavior is not revealed by Algorithms 3 and 4, which yield a monotonic decreasing velocity curve. The non monotonic trend of velocity for increasing concentrations of nocodazole is significant from a biological standpoint, since it was hypothesized but rarely verified in practice, and might help to better dose nocodazole in chemotherapeutic treatments.

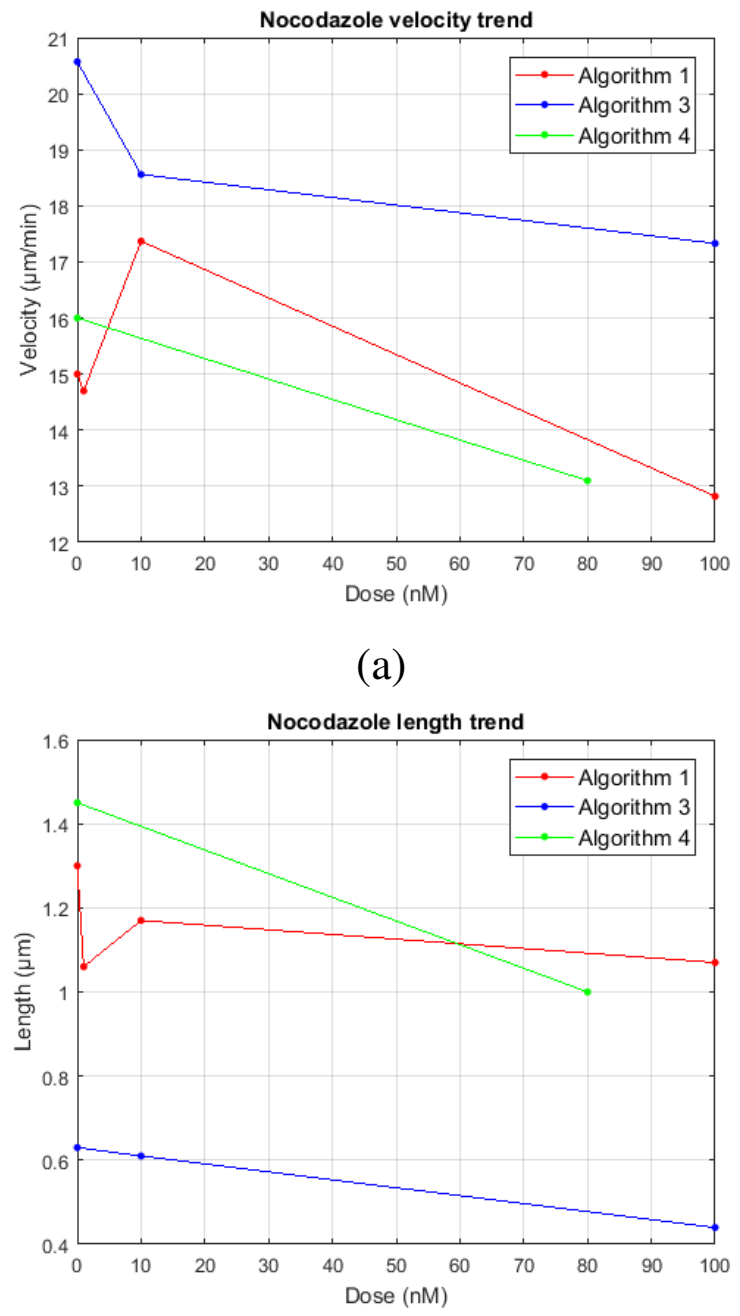

(a) (b)

Fig. 6: Nocodazole-doped cultures: comparison between Algorithms 1, 3 and 4

Finally, the results of our algorithm have been compared with those computed by hand by expert biologists of the Department of Molecular Biotechnology and Health Sciences of the University of Turin. This has been done on a subset of the same data stacks, referring to nocodazole-doped cell cultures. The available comparisons are listed below in Table VIII; only mean velocity values have been taken into account.

| NOCODAZOLE | | | |
|---|---|---|---|
| Dose | Stack ID | Manual | Algo 1 |
| 0 nM | 003 | 12.03 | 12.24 |
| | 005 | 16.86 | 15.36 |
| | 011 | 16.77 | 15.00 |
| 1 nM | 002 | 14.02 | 13.34 |
| | 005 | 23.53 | 14.00 |
| | 007 | 16.60 | 16.92 |

TABLE VIII: Comparison with manually scored data (nocodazole-doped cells)

It is worth pointing out that at 1 nM concentration of nocodazole, only tracks longer than $1\mu m$ have been taken into account. The comparison with manual screeing provides encouraging results. All values exhibits non-significant difference, with the exception of series 005 of nocodazole at 1 nM concentration. However, that value can be regarded as an outlier, since it is an unusual velocity value which is not found in the other cells doped with the same drug.

## VII. Conclusions

This work aims to supply an automatic tool for tracking and analyzing astral microtubule behavior in fluorescence images. Despite the lack of a generally accepted ground truth, the validation process has provided encouraging results, which are also well-substantiated by the expected drug effects that can be found in literature. An important matter to highlight is the benefit in terms of computational time; indeed the time spent on analyzing samples through the proposed algorithm, are considerably less, if compared to the manual labor (we are talking about several hours compared to few minutes). Moreover, since this is an automatic software, it is not affected by human errors, due to tiredness or attention deficit, that unavoidably arise from such challenging work. Indeed, it is not uncommon that errors are committed when the analysis results are transcribed, as the gap found between manual and automatic evaluation (i.e. series 005 of nocodazole at 1 nM) could demonstrate. This is the reason why the developed algorithm can provide a support for manual experiments, without completely replace it. It is worth pointing out that, with this in mind, the proposed software is currently being tested by the biologists of the Department of Molecular Biotechnology and HealthSciences of University of Turin; in this way other samples, different from those employed in this thesis, can be evaluated. As regards future work, the algorithm might be improved with a better management of the denoising stage; moreover, in order to improve performance, detection thresholds should be adapted to each stack. Thus, problems related to the false positive rate could be further limited. Finally, to ensure a better portability, it is planned to leave the MathWorks environment developing an ImageJ plug-in.